\documentclass[12pt]{article}

\usepackage{amsmath}
\usepackage{amssymb}
\usepackage{mathtools}
\newcommand{\be}{\begin{equation}}
\newcommand{\bea}{\begin{eqnarray}}
\newcommand{\ee}{\end{equation}}
\newcommand{\eea}{\end{eqnarray}}

\allowdisplaybreaks[1]
\begin{document}

\renewcommand{\thefootnote}{\fnsymbol{footnote}}
\vskip .4in
\begin{center}
{\Large \bf Quartic interaction vertex in the massive integer higher
spin field theory in a constant electromagnetic field

 } \vskip .4in {\large I.L. Buchbinder$^{ab}$\footnote{e-mail: {\tt joseph@tspu.edu.ru}},
V.A.Krykhtin$^{a}$\footnote{e-mail: {\tt  krykhtin@tspu.edu.ru }} }
\vskip .2in {$^a$ \it Department of Theoretical Physics, Tomsk State Pedagogical University,\\
Tomsk, 634061, Russia} \\
\vskip .1in {\it ${}^b$National Research Tomsk State University,
Tomsk, 634050, Russia}

\begin{abstract}
We consider the massive integer higher spin fields coupled to an
external constant electromagnetic field in flat space of arbitrary
dimension and find a gauge invariant quartic interaction vertex
which is quadratic in dynamical higher spin field and quadratic in
external field. Construction of the vertex is based on the BRST
approach to higher spin filed theory where no off-shell constraints
on the fields and on the gauge parameters are imposed from the very
beginning (unconstrained formulation).
\end{abstract}

\end{center}

\renewcommand{\thefootnote}{\arabic{footnote}}
\setcounter{footnote}{0}

\section{Introduction}

The higher spin field theory is the actively developing trend in
modern theoretical and mathematical physics (see e.g. the reviews
\cite{Reviews}). One of the interesting and important problems here
is related with construction of the Lagrangians describing the
interactions of the higher spin fields among themselves and with the
various backgrounds (see e.g. the papers
\cite{interactions1,interactions2,Klishevich:1998ng,interactions3}
and the references therein).

Most of the works on interacting Lagrangians of the higher spin
fields deal with the cubic approximation. In particular, the cubic
vertices of higher spin fields have been constructed in
\cite{interactions1}, the cubic vertices for higher spin fields
coupled to external electromagnetic and gravitational backgrounds
have been found in \cite{interactions2}. The quartic approximation
of the interaction vertices has been studied much less (see e.g.
\cite{Klishevich:1998ng,interactions3}).

When considering interactions of higher spin fields with a
nontrivial background one faces several difficulties such as a
possibility of superluminal propagation and violation of the number
of physical degrees of freedom. The requirement that the
superluminal propagation is absent imposes in general the certain
conditions on the background fields \cite{Velo:1969bt} (see also
\cite{BKL} for a recent discussion). Similarly, when turning on
nonzero background fields the invariance of the initial system under
its gauge transformations can be partially or completely lost and
this means in turn that nonphysical polarizations can appear in the
spectrum. The requirement of preserving the physical degrees of
freedom generically imposes some extra conditions on the background.
Therefore to construct a higher spin field theory on some background
we should answer the question whether the background under
consideration is physically acceptable i.e., whether it satisfies
the constraints imposed by the above mentioned conditions.

In this paper we study a problem of interaction of massive totally
symmetric bosonic higher spin fields with constant electromagnetic
(EM) background in Minkowski space  of an arbitrary dimension $d$.
These higher spin fields are described by tensors with an arbitrary
number $s$ of totally symmetric tensorial indices. Our main aim is
to derive the gauge invariant Lagrangian using the method of BRST
construction \cite{BRST} in the quadratic approximation in strength
$F_{\mu\nu}$ of the external field. This method in fact yields a
gauge invariant Lagrangian description for massive higher spin
fields in extended Fock space and therefore the Lagrangian will
contain, apart from the basic fields, some additional fields such as
St\"{u}ckelberg fields. Some of these fields are eliminated with the
help of gauge transformations, some of the others should be
eliminated as a result of the equations of motion. Therefore, in
order to have a consistent gauge invariant description for massive
higher spin fields, one should have enough gauge freedom and have
the ``correct" equations of motion, which ensure the absence of
ghosts\footnote{One way to check this is to perform a complete gauge
fixing in the equations of motion and obtain the equations in terms
of basic fields. As a result one obtains equations defining the
spectrum of the theory and check if it is ghost free or not. }. One
can  show that the preservation of physical degrees of freedom
indeed takes place for the Lagrangian under consideration, provided
that the terms containing the strength of the external field are
considered as a perturbation.

The paper is organized as follows.  Section~\ref{Fne0} contains our
main results. After a brief reminder of construction of Lagrangians
for free massive bosonic higher spin fields we introduce interaction
with background electromagnetic fields by deforming the operators
which define the BRST charge. The requirement that the deformed
operators form a closed algebra fixes the free parameters which are
present in the definition of the deformed operators. Then we
construct the corresponding BRST charge, derive the gauge invariant
Lagrangian describing the interactions with a constant
electromagnetic background. In Section~\ref{secquartet} we derive
from our main Lagrangian, obtained in Section~\ref{Fne0}, 
a family of the Lagrangian formulations with different sets of the
auxiliary fields by partially or completely gauge fixing
As examples we obtain first the Lagrangian
description in terms of so called ``quartet formulation"
\cite{Buchbinder:2007ak,Buchbinder:2008ss} (see also
\cite{Pashnev:1989gm}) and second the Lagrangian formulation
obtained in \cite{Klishevich:1998ng}. Actually the expressions for
the operators obtained in Section~\ref{Fne0} from the very beginning
are more general in compare with earlier work
\cite{Klishevich:1998ng} and hence we obtain new interaction
vertices for quartic coupling of higher spin fields to constant
electromagnetic background. The final Section contains the
conclusions and a discussions.

\section{BRST Lagrangian construction for massive higher spin bosonic fields on electromagnetic background} \label{Fne0}
\setcounter{equation}0 Let us briefly summarize the features of the
BRST approach for the construction of the gauge invariant bosonic
and fermionic higher spin field Lagrangians \cite{BRST}. First one
introduces a set of operators that define a spectrum of the
theory\footnote{In free theory the spectrum is given with the help
of the relations defining either reducible or  irreducible
representations of the Poincare or AdS group.}. Provided these
operators form a closed algebra one builds a nilpotent BRST charge
$Q$, which in turn yields a quadratic gauge invariant Lagrangian in
terms of the Fock sate vectors of the form
\begin{equation}
{\cal  L}  \sim \langle \chi |Q| \chi \rangle
\end{equation}
where $| \chi \rangle$ is a vector in an extended Fock space. The
gauge invariance of the Lagrangian under the linear gauge
transformations
\begin{equation}\label{L}
 \delta| \chi \rangle = Q| \Lambda \rangle
\end{equation}
is guaranteed by the nilpotency of the BRST charge, $Q^2=0$.

The situation can be  more complicated if the closure of the algebra
of the initial set of operators requires inclusion of certain
additional operators into the above set. These extra operators can
impose too strong conditions on the field $|\chi \rangle$ so that
there will be no nonzero solution to the equation
\begin{equation} \label{Q}
Q|\chi\rangle=0.
\end{equation}
This problem is overcome as follows \cite{BRST}. One introduces the
additional sets of  oscillator variables and builds an auxiliary
representation of the generators of the algebra (i.e. of the
operators under consideration) in terms of these new variables. Then
one defines a modified set of operators as a sum of new and old ones
and considers the problem in an extended Fock space (including the
new oscillators). After that one builds BRST charge for modified
generators in the standard way since the generators form a closed
algebra. It allows us to construct a Lagrangian on the base of the
BRST charge under consideration.

After this reminder let us turn  to a description of massive
bosonic  higher spin fields. To this end we introduce the  Fock
space spanned by the oscillators
\begin{equation}
[a_\mu,a_\nu^+]=\eta_{\mu\nu},, \quad \eta_{\mu\nu}= diag (-1,1,...,1)
\end{equation}
and consider the  operators
\begin{equation} \label{properators}
l^\prime_0=\partial^2-m^2, \qquad l^\prime_1=
ia^\mu\partial_\mu,
\qquad  l_2^\prime=\tfrac{1}{2}a^{\mu}a_\mu
\quad.
\end{equation}
The first of the operators in (\ref{properators}) corresponds to the
Klein-Gordon operator for the massive boson, the second operator is a divergence
operator, and the third one is an operator which takes a trace. In order to
have a hermitian BRST charge we also introduce  operators which
are hermitian conjugate to the operators $l^\prime_1$
and $ l_2^\prime$
\begin{equation}
l^{\prime +}_1=ia^{+\mu }\partial_\mu,
\qquad  l_2^{\prime +}=\tfrac{1}{2}a^{+\mu }a^+_\mu
.
\end{equation}
Finally in order to close the algebra one introduces the extra
operators
\begin{equation}
 g'_0=a^{+ \mu} a_\mu+\frac{d}{2}
\end{equation}
and $g'_{m}=m^2$. The operator $g'_0$ is a ``particle" number
operator and its eigenvalues are  always strictly positive.
Therefore, we have a situation described earlier in this Section. We
introduce two sets of additional bosonic oscillator variables with commutation relations
\begin{equation} \label{b12}
[b_1,b_1^+]=1, \quad
[b_2,b_2^+]=1.
\end{equation}
Using these new variables one can build auxiliary representation for
the original operators and define modified operators as
\begin{align}
\label{op-0}
&
l_0=\partial^2-m^2
\\
&
l_1=ia^\mu\partial_\mu+mb_1
&&
l_1^+=ia^{+\mu}\partial_\mu+mb_1^+
\\
&
\label{op-l2}
l_2=\tfrac{1}{2}a^{\mu}a_\mu+\tfrac{1}{2}b_1^2
+(b_2^+b_2+h)b_2
&&
l_2^+=\tfrac{1}{2}a^{+\mu}a_\mu^++\tfrac{1}{2}b_1^{+2}+b_2^+
\\
& g_0=a^+_\mu a^\mu+b_1^+b_1+2b_2^+b_2+\tfrac{d+1}{2}+h
&&
g_{m}=0
\label{op-g0}
\end{align}
where $h$ is an arbitrary real constant. The algebra of these
operators is given by Table 1.

In order to introduce an interaction of the bosonic fields with an
external constant EM background  field $F_{\mu\nu}=const$ we shall
proceed as follows. First we replace all the partial derivatives by
the $U(1)$ covariant ones $D_\mu=\partial_\mu-ieA_\mu$ and include
into the expressions of the operators\footnote{We shall denote these
new operators by the corresponding capital letters.}
(\ref{op-0})--(\ref{op-g0})  terms which  vanish in the limit
$F_{\mu\nu}\to0$. After that we require that the new  operators form
a closed algebra.

Before writing an ansatz for the operators let us note that since
the trace of a field and its traceless part are independent from
each  other one can shift the trace of a field so that the traceless
condition remains unchanged. Thus we suppose that the operators
related with the traceless condition $l_2$, $l_2^+$
as well as the number operator $g_0$  remain unchanged
\begin{align}
&L_2=l_2, &&L_2^+=l_2^+&&G_0=g_0.
\label{T1}
\end{align}
Moreover, since the oscillator variables $b_2$, $b_2^+$
(\ref{b12}) are included only in operators (\ref{T1})
(see also the expressions (\ref{op-l2})--(\ref{op-g0})) we  assume
that these variables  are not present in the expressions of the
operators $L_0$, $L_1$, $L_1^+$.

Since we are going to consider first the linear in $F_{\mu\nu}$ approximation we take the following ansatz for the operators
\begin{eqnarray}
L_1
&=&
ia^\alpha D_\alpha+mb_1
+\frac{ie}{m}a^{+\mu}F_{\mu\alpha}a^\alpha\sum_{k=0}^\infty\alpha_{0(k)}(b_1^+)^k(b_1)^{k+1}
\nonumber
\\*
&&{}
+\frac{e}{m^2}a^\alpha F_{\alpha\sigma}D^\sigma\sum_{k=0}^\infty\beta_{0(k)}(b_1^+)^k(b_1)^k
+\frac{e}{m^2}a^{+\mu}F_{\mu\sigma} D^\sigma\sum_{k=0}^\infty\gamma_{0(k)}(b_1^+)^k(b_1)^{k+2}
\label{op-L1}
\\
L_0
&=&
D^2-m^2
+iea^{+\mu}F_{\mu\alpha}a^\alpha\sum_{k=0}^\infty\xi_{0(k)}(b_1^+)^k(b_1)^k
\nonumber
\\*
&&{}
+\frac{e}{m}a^{+\mu}F_{\mu\sigma}D^\sigma\sum_{k=0}^\infty\zeta_{0(k)}(b_1^+)^k(b_1)^{k+1}
+\frac{e}{m}a^\alpha F_{\alpha\sigma}D^\sigma\sum_{k=0}^\infty\zeta_{0(k)}'(b_1^+)^{k+1}(b_1)^{k}
\\
L_1^+
&=&
ia^{+\mu}D_\mu +mb_1^+
+\frac{ie}{m}a^{+\mu}F_{\mu\alpha}a^\alpha\sum_{k=0}^\infty\alpha_{0(k)}'(b_1^+)^{k+1}(b_1)^{k}
\nonumber
\\*
&&{}
+\frac{e}{m^2}a^{+\mu}
  F_{\mu\sigma}D^\sigma\sum_{k=0}^\infty\beta_{0(k)}'(b_1^+)^k(b_1)^k
+\frac{e}{m^2}a^\alpha
F_{\alpha\sigma}D^\sigma\sum_{k=0}^\infty\gamma_{0(k)}'(b_1^+)^{k+2}(b_1)^k
\label{op-L1+}
\end{eqnarray}
where $\alpha_{0(k)}$, \dots, $\gamma_{0(k)}'$ are arbitrary complex
constants and the  rest of the operators (\ref{op-l2})--(\ref{op-g0}) are
unchanged as one can see form the equation (\ref{T1}).
Let us note that the above relations can be treated as
the deformations of the corresponding relations of free theory by
the terms linear in $F_{\mu\nu}$.

Let us point out that the ansatz for the operators $L_1$, $L_0$,
$L_1^+$ (\ref{op-L1})--(\ref{op-L1+}) is not the most general one. The
ansatz is taken on the basis  of the following ``minimal" rule. Let us
consider the operators (\ref{op-L1})--(\ref{op-L1+}) in free theory,
replace the partial derivatives by the covariant ones and calculate
the commutators. Obviously the  algebra will not be
closed. Then one adds to these operators the minimal number of terms
linear in $F_{\mu\nu}$ in such a way that  the  algebra is closed in the
linear approximation. One can see that according to this ``minimal"
rule the Lorentz indices of the creation and annihilation operators
are always contracted with an index or indices of $F_{\mu\nu}$.
In principle it is possible to consider  other deformations of the
free theory by the terms linear in $F_{\mu\nu}$. For example, one can
add to $L_1$ a term of the form
$a_\mu^+a^\mu a^\alpha F_{\alpha\sigma}D^\sigma$
but this term  does not obey the ``minimal" rule.

From the requirement the $L_0$ to be  hermitian, from the
condition $(L_1)^+=L_1^+$ and from the requirement that the total
system of operators forms a closed algebra in the linear
approximation one finds the expressions for constants which are
present in (\ref{op-L1})--(\ref{op-L1+})
and these constants are expressed through four arbitrary real constants $c_1$, $c_2$, $c_3$, $c_4$.
These expressions are summarized in the Appendix.

Let us consider the second order approximation in $F_{\mu\nu}$.
Ansatz for corrections to operators $L_1$, $L_0$ and $L_1^+$ in this case looks like
\begin{eqnarray}
L_1^{(2)}
&=&
\frac{e^2}{m^3}F^2\sum_{k=0}^\infty\alpha_{1k}(b_1^+)^k(b_1)^{k+1}
+\frac{e^2}{m^3}a^{+\mu}F^2_{\mu\alpha}a^{\alpha}\sum_{k=0}^\infty\alpha_{2k}(b_1^+)^k(b_1)^{k+1}
\nonumber
\\*
&&
{}
+\frac{e^2}{m^3}(a^{+\mu}F_{\mu\alpha}a^{\alpha})^2\sum_{k=0}^\infty\alpha_{3k}(b_1^+)^k(b_1)^{k+1}
\nonumber
\\*
&&
{}
+\frac{e^2}{m^3}F^2_{\alpha\beta}a^{\alpha\beta}\sum_{k=0}^\infty\alpha_{4k}(b_1^+)^{k+1}(b_1)^k
+\frac{e^2}{m^3}a^{+\mu\nu}F^2_{\mu\nu}\sum_{k=0}^\infty\alpha_{5k}(b_1^+)^k(b_1)^{k+3}
\nonumber
\\*
&&
{}
+\frac{ie^2}{m^4}a^{\alpha}F^2_{\alpha\sigma}D^\sigma\sum_{k=0}^\infty\beta_{1k}(b_1^+)^{k}(b_1)^k
+\frac{ie^2}{m^4}a^{+\mu}F^2_{\mu\sigma}D^\sigma\sum_{k=0}^\infty\gamma_{1k}(b_1^+)^k(b_1)^{k+2}
\nonumber
\\*
&&
{}
+\frac{ie^2}{m^4}(a^{+\mu}F_{\mu\alpha}a^{\alpha})a^{\alpha}F_{\alpha\sigma}D^\sigma
\sum_{k=0}^\infty\beta_{2k}(b_1^+)^{k}(b_1)^k
\nonumber
\\*
&&
{}
+\frac{ie^2}{m^4}a^{+\mu}F_{\mu\sigma}D^\sigma(a^{+\mu}F_{\mu\alpha}a^{\alpha})
\sum_{k=0}^\infty\gamma_{2k}(b_1^+)^k(b_1)^{k+2}
\label{op-L12}
\\
L_0^{(2)}
&=&
\frac{e^2}{m^2}F^2\sum_{k=0}^\infty\xi_{1k}(b_1^+)^k(b_1)^k
+\frac{e^2}{m^2}a^{+\mu}F^2_{\mu\alpha}a^{\alpha}\sum_{k=0}^\infty\xi_{2k}(b_1^+)^k(b_1)^k
\nonumber
\\*
&&{}
+\frac{e^2}{m^2}(a^{+\mu}F_{\mu\alpha}a^{\alpha})^2\sum_{k=0}^\infty\xi_{3k}(b_1^+)^k(b_1)^k
+\frac{e^2}{m^2}F^2_{\alpha\beta}a^{\alpha\beta}\sum_{k=0}^\infty\xi_{4k}(b_1^+)^{k+2}(b_1)^k
\nonumber
\\*
&&
{}
+\frac{e^2}{m^2}a^{+\mu\nu}F^2_{\mu\nu}\sum_{k=0}^\infty\xi_{4k}'(b_1^+)^k(b_1)^{k+2}
+\frac{ie^2}{m^3}a^{\alpha}F^2_{\alpha\sigma}D^\sigma\sum_{k=0}^\infty\zeta_{1k}(b_1^+)^{k+1}(b_1)^k
\nonumber
\\*
&&
{}
+\frac{ie^2}{m^3}a^{+\mu}F^2_{\mu\sigma}D^\sigma\sum_{k=0}^\infty\zeta_{1k}'(b_1^+)^k(b_1)^{k+1}
+\frac{ie^2}{m^3}(a^{+\mu}F_{\mu\alpha}a^{\alpha})a^{\alpha}F_{\alpha\sigma}D^\sigma
\sum_{k=0}^\infty\zeta_{2k}(b_1^+)^{k+1}(b_1)^k
\nonumber
\\*
&&{}
+\frac{ie^2}{m^3}a^{+\mu}F_{\mu\sigma}D^\sigma(a^{+\mu}F_{\mu\alpha}a^{\alpha})
\sum_{k=0}^\infty\zeta_{2k}'(b_1^+)^k(b_1)^{k+1}
\nonumber
\\*
&&
{}
+\frac{e^2}{m^4}F^2_{\sigma\tau}D^\sigma D^\tau\sum_{k=0}^\infty\zeta_{3k}(b_1^+)^{k}(b_1)^k
+\frac{e^2}{m^4}(a^{+\mu}F_{\mu\sigma}D^\sigma)(a^{\alpha}F_{\alpha\sigma}D^\sigma)
\sum_{k=0}^\infty\zeta_{4k}(b_1^+)^{k}(b_1)^k
\nonumber
\\*
&&
{}
+\frac{e^2}{m^4}(a^{\alpha}F_{\alpha\sigma}D^\sigma)^2\sum_{k=0}^\infty\zeta_{5k}(b_1^+)^{k+2}(b_1)^k
+\frac{e^2}{m^4}(a^{+\mu}F_{\mu\sigma}D^\sigma)^2\sum_{k=0}^\infty\zeta_{5k}'(b_1^+)^k(b_1)^{k+2}
\\
L_1^{+(2)}
&=&
\frac{e^2}{m^3}F^2\sum_{k=0}^\infty\alpha_{1k}'(b_1^+)^{k+1}(b_1)^{k}
+\frac{e^2}{m^3}a^{+\mu}F^2_{\mu\alpha}a^{\alpha}\sum_{k=0}^\infty\alpha_{2k}'(b_1^+)^{k+1}(b_1)^{k}
\nonumber
\\*
&&
{}
+\frac{e^2}{m^3}(a^{+\mu}F_{\mu\alpha}a^{\alpha})^2\sum_{k=0}^\infty\alpha_{3k}'(b_1^+)^{k+1}(b_1)^{k}
\nonumber
\\
&&
{}
+\frac{e^2}{m^3}a^{+\mu\nu}F^2_{\mu\nu}\sum_{k=0}^\infty\alpha_{4k}'(b_1^+)^k(b_1)^{k+1}
+\frac{e^2}{m^3}F^2_{\alpha\beta}a^{\alpha\beta}\sum_{k=0}^\infty\alpha_{5k}'(b_1^+)^{k+3}(b_1)^k
\nonumber
\\*
&&
{}
+\frac{ie^2}{m^4}a^{+\mu}F^2_{\mu\sigma}D^\sigma\sum_{k=0}^\infty\beta_{1k}'(b_1^+)^k(b_1)^{k}
+\frac{ie^2}{m^4}a^{\alpha}F^2_{\alpha\sigma}D^\sigma\sum_{k=0}^\infty\gamma_{1k}'(b_1^+)^{k+2}(b_1)^k
\nonumber
\\*
&&
{}
+\frac{ie^2}{m^4}a^{+\nu}F_{\nu\sigma}D^\sigma(a^{+\mu}F_{\mu\alpha}a^{\alpha})
\sum_{k=0}^\infty\beta_{2k}'(b_1^+)^k(b_1)^{k}
\label{op-L1+2}
\end{eqnarray}
where $\alpha_{i(k)}$, \dots, $\gamma_{j(k)}'$ are arbitrary complex
constants.

Analogously to the linear approximation from the requirement the
$L_0$ to be  hermitian, from the condition $(L_1)^+=L_1^+$ and from
the requirement that the total system of operators forms a closed
algebra in the quadratic approximation one finds the expressions for
constants which are present in (\ref{op-L12})--(\ref{op-L1+2}).
These expressions are summarized in the Appendix and all of them are
expressed through eight independent real constants. One should note
that the quadratic approximation gives the additional restrictions
on the arbitrary real constants $c_1$ and $c_3$ of the linear
approximation (\ref{c1c3}).

Note that  a similar problem was considered in
\cite{Klishevich:1998ng}, but our approach leads to more general
Lagrangian construction. We found two more arbitrary constants in
the linear approximation and six\footnote{Two of the four arbitrary
constants in the quadratic approximation in \cite{Klishevich:1998ng}
are related with deformation of the trace condition.} more arbitrary
constants in the quadratic approximation because, unlike
\cite{Klishevich:1998ng}, we do not require from the very beginning
that the coefficients in (\ref{op-L1})--(\ref{op-L1+}) and
(\ref{op-L12})--(\ref{op-L1+2}) must satisfy reality conditions. As
one can see from the Appendix, the complex coefficients are also
acceptable.

The new operators form an algebra which is the same as in the free
theory and it is given in Table \ref{table0}.
\begin{table}[t]
\small
\begin{eqnarray*}
\begin{array}{||c||c|c|c|c|c||c||}\hline\hline\vphantom{\biggm|}\hspace{-0.3em}
[\;\downarrow\;,\to\}\hspace{-0.4em}&\quad L_0&L_1&L_1^+&L_2&L_2^+ &G_0\\
\hline\hline\vphantom{\biggm|}
L_0
   &0&0&0&0&0&0\\
\hline\vphantom{\biggm|}
L_1
  &0&0&-L_0&0&L_1^+&L_1 \\
\hline\vphantom{\biggm|}
L_1^+
   &0&L_0&0&-L_1&0&-L_1^+\\
\hline\vphantom{\biggm|}
L_2
   &0&0&L_1&0&G_0&2L_2\\
\hline\vphantom{\biggm|}
L_2^+
   &0&-L_1^+&0&-G_0&0&-2L_2^+\\
\hline\hline\vphantom{\biggm|}
G_0
   &0&-L_1&L_1^+&-2L_2&2L_2^+&0\\
\hline\hline
\end{array}
\end{eqnarray*}
\caption{The algebra of the operators.}\label{table0}
\end{table}
After we have achieved the closure of the algebra for the operators, the next step is to construct the
corresponding BRST charge.
This procedure follows closely the one developed for the bosonic fields in
\cite{BRST}
to which we refer for more details.
First we construct the standard BRST operator on the basis of the operators (\ref{T1})--(\ref{op-L1+2})
\begin{eqnarray}\label{QQ}
Q&=&
\eta_0L_{0}+\eta_1^+L_1+\eta_1L_1^++\eta_2^+L_2+\eta_2L_2^++\eta_{G}G_0
+\eta_1^+\eta_1{\cal{}P}_0
-\eta_2^+\eta_2{\cal{}P}_G
\nonumber
\\&&
{}
+(\eta_G\eta_1^+-\eta_2^+\eta_1){\cal{}P}_1
+(\eta_1\eta_G-\eta_1^+\eta_2){\cal{}P}_1^+
+2\eta_G\eta_2^+{\cal{}P}_2
+2\eta_2\eta_G{\cal{}P}_2^+.
\end{eqnarray}
Here $\eta_0$, $\eta_1^+$, $\eta_1$, $\eta_2^+$, $\eta_2$, $\eta_G$ are fermionic ghost ``coordinates'' corresponding to their canonically conjugate ghost ``momenta'' ${\cal{}P}_0$, ${\cal{}P}_1$, ${\cal{}P}_1^+$, ${\cal{}P}_2$, ${\cal{}P}_2^+$, ${\cal{}P}_G$. They obey the anticommutation relations
\begin{eqnarray}
\label{ghosts}
&
\{\eta_1,{\cal{}P}_1^+\}= \{{\cal{}P}_1, \eta_1^+\}
=
\{\eta_2,{\cal{}P}_2^+\}= \{{\cal{}P}_2, \eta_2^+\}
=\{\eta_0,{\cal{}P}_0\}= \{\eta_G,{\cal{}P}_G\} =
1
\end{eqnarray}
and possess the standard  ghost number distribution,
$gh(\eta)=-gh(\mathcal{P})=1$,
which gives    $gh({Q})=1$.

For the subsequent computations it is convenient to present the  BRST operator (\ref{QQ})
in the form
\begin{eqnarray*}
Q&=&
\tilde{Q}+\eta_G\bigl(N+\tfrac{d-5}{2}+h\bigr)+(2q_1^+q_1-\eta_2^+\eta_2){\cal{}P}_G
\nonumber
\\[0.5em]
{}
N&=&a^+_\mu a^\mu+b_1^+b_1+2b_2^+b_2
+\eta_1^+{\cal{}P}_1+{\cal{}P}_1^+\eta_1+2\eta_2^+{\cal{}P}_2+2{\cal{}P}_2^+\eta_2
\\
[0.5em]
\tilde{Q}&=&
\eta_0L_0+\Delta Q+\eta_1^+\eta_1{\cal{}P}_0
\\
[0.5em]
\Delta Q&
=&
\eta_1^+L_1+\eta_1L_1^++\eta_2^+L_2+\eta_2L_2^+
-\eta_2^+\eta_1{\cal{}P}_1
-\eta_1^+\eta_2{\cal{}P}_1^+
\end{eqnarray*}
Next  we choose the following representation for the vacuum in the
Fock space
\begin{equation}
\left(\mathcal{P}_0,{\cal{}P}_G, \eta_1, {\cal{}P}_1, \eta_2,{\cal{}P}_2\right)|0\rangle =0\,,
\label{ghostvac}
\end{equation}
and suppose
that the  vectors and gauge parameters do not depend on  $\eta_G$,
\begin{eqnarray}\label{chi}
 |\chi\rangle &=&\sum_{k_i}
(\eta_0)^{k_1}(\eta_1^+)^{k_2}(\mathcal{P}_1^+)^{k_3}(\eta_2^+)^{k_4} (\mathcal{P}_2^+)^{k_5}
  (b_1^+)^{k_{6}}(b_2^+)^{k_{7}}
a^{+{}\mu_1}\cdots a^{+{}\mu_{k_0}}\chi^{k_1 \cdots k_{7}}_{\mu_1\cdots \mu_{k_0}}(x)
   |0\rangle.
\end{eqnarray}
The sum in (\ref{chi}) is taken  over $k_0, k_6, k_7$, running from 0 to infinity,
and over $k_1$, $k_2,$ $k_3,$ $k_4,$ $k_5$, running from 0 to 1.
Then, we derive from the
equations (\ref{Q}) as well as from the reducible gauge transformations, (\ref{L}) a sequence of relations
\begin{align}
& \tilde{Q}|\chi\rangle=0, && (N+\tfrac{d-5}{2}+h)|\chi\rangle=0, &&
\left(\epsilon, {gh}\right)(|\chi\rangle)=(1,0),
\label{Qchi}
\\
& \delta|\chi\rangle=\tilde{Q}|\Lambda\rangle, &&
(N+\tfrac{d-5}{2}+h)|\Lambda\rangle=0, && \left(\epsilon,
{gh}\right)(|\Lambda\rangle)=(0,-1),
\label{QLambda}
\\
& \delta|\Lambda\rangle=\tilde{Q}|\Lambda^{(1)}\rangle, &&
(N+\tfrac{d-5}{2}+h)|\Lambda^{(1)}\rangle=0, && \left(\epsilon,
{gh}\right)(|\Lambda^{(1)}\rangle)=(1,-2).
\label{QLambdai}
\end{align}
Here $\epsilon$ defines a Grassmann parity of  corresponding fields and parameters of gauge transformations
as $(-1)^\epsilon$.
The middle equation in (\ref{Qchi}) is a constraint on   possible values of $h$
\begin{eqnarray}
\label{h}
h=-s-\frac{d-5}{2}.
\end{eqnarray}
By fixing the value of spin, we also fix the parameter $h$, according to
(\ref{h}).
Having fixed a value of $h$, we then  substitute it into each of
the expressions (\ref{Qchi})--(\ref{QLambdai}).

It is straightforward to check that the equations (\ref{Qchi}) can be obtained from the following Lagrangian
\cite{BRST}
\begin{eqnarray}
\mathcal{L}&=&
\int d\eta_0\,\, \langle\chi|K_h\tilde{Q}|\chi\rangle
\nonumber
\\*
&=&
\langle{}S|K_hL_0|S\rangle-\langle{}S|K_h\Delta{}Q|A\rangle
-
\langle{}A|K_h\Delta{}Q|S\rangle-\langle{}A|K_h\eta_1^+\eta_1|A\rangle
,
\label{LagrF}
\end{eqnarray}
where we have decomposed state vector $|\chi\rangle$ as follows
\begin{eqnarray}
|\chi\rangle=|S\rangle+\eta_0|A\rangle.
\label{SA}
\end{eqnarray}
In (\ref{LagrF}) operator $K_h$
\begin{eqnarray}
K_h&=&\sum_{n=0}^\infty
     |n\rangle
     \frac{C(n,h)}{n!}
     \langle{}n|,
\qquad
|n\rangle=(b_2^+)^n|0\rangle,
\qquad
h+s+\frac{d-5}{2}=0,
\\
&&
C(n,h)=h(h+1)(h+2)\ldots(h+n-1),
\qquad
\qquad
C(0,h)=1.
\end{eqnarray}
is
needed to maintain Hermiticity of the Lagrangian since
 as one can see from the auxiliary representations for operators (\ref{op-l2}) one has
$(l_2)^+ \neq l_2^+$.

The Lagrangian (\ref{LagrF}) describes  the interaction of massive bosonic fields with constant electromagnetic field in quadratic approximation in $F_{\mu\nu}$ and it is  our main result.
It contains, apart from the basic field $\varphi_{\mu_1\ldots\mu_s}(x)$ in $|S\rangle$ (\ref{SA})
\begin{eqnarray}
|S\rangle=\varphi_{\mu_1\ldots\mu_n}(x)a^{+\mu_1}\ldots a^{+\mu_n}|0\rangle
+\ldots
\end{eqnarray}
a number of additional fields\footnote{In decomposition (\ref{chi}) they are coefficients in summands which contain at least one creation operator different from $a^{+\mu}$.}, whose number increases with spin value.

In the next section we show that after  partially or completely gauge fixing
one can obtain different Lagrangian formulations with  a smaller number of additional fields.

\setcounter{equation}0\section{Partial gauge fixing and different Lagrangian formulations} \label{secquartet}

In this Section we are going to obtain from (\ref{LagrF}) the
various equivalent Lagrangian formulations by partially fixing the
gauge invariance.

First we derive a quartet Lagrangian formulation
\cite{Buchbinder:2007ak,Buchbinder:2008ss}. Initially this
formulation was developed for the massless higher spin fields in
flat and AdS background in \cite{Buchbinder:2007ak}. Its bosonic
version  contains six unconstrained fields (one physical field and
five additional fields two of which are Lagrangian multipliers) and
one unconstrained gauge parameter.\footnote{Another similar
formulation (so-called triplet formulation) of bosonic fields on
Minkowski and $AdS_d$ backgrounds contains one physical and two
additional fields \cite{Francia:2002pt} (see also
\cite{Bekaert:2015fwa} for a recent discussion) and corresponds to a
description of reducible representations of the Poincare or
$SO(d-1,2)$ groups.} Using  dimensional reduction one can obtain in
principle the quartet formulation for massive higher spin fields
\cite{Buchbinder:2008ss}.

To derive the unconstrained formulation from the Lagrangian
(\ref{LagrF}) we partially fix gauge invariance as it was done in
\cite{BRST} (putting there the curvature to zero
$R=0$), except we will not fix gauge invariance corresponding to
gauge parameter $|\varepsilon\rangle$ (we remove dependence only on
$b_2^+$)
\begin{eqnarray}
|\Lambda\rangle
&=&
\mathcal{P}_1^+|\varepsilon\rangle
+\ldots,
\\
|\varepsilon\rangle&=&\sum_{k=0}^{s-1}\frac{1}{k!}(b_1^+)^k|\varepsilon_{n-k-1}\rangle,
\qquad
|\varepsilon_{s-k-1}\rangle=\frac{1}{(s-k)!}\;a^{+\mu_1}\ldots a^{+\mu_{s-k-1}}\varepsilon_{\mu_1\ldots\mu_{s-k-1}}(x)|0\rangle
\label{decomp-q}
.
\end{eqnarray}
Next one can show that after the gauge fixing some of the remaining
fields can be removed with the help of the equations of motion
analogously to \cite{Buchbinder:2007ak} and the non-vanishing fields
in the quartet formulation are
\begin{eqnarray}
|S\rangle&=&|S_1\rangle+\eta_1^+\mathcal{P}_1^+|S_2\rangle+\eta_2^+\mathcal{P}_1^+|S_4\rangle
\label{decS}
\\
|A\rangle&=&\mathcal{P}_1^+|A_1\rangle+\mathcal{P}_2^+|A_2\rangle+\eta_1^+\mathcal{P}_1^+\mathcal{P}_2^+|A_3\rangle.
\label{decA}
\end{eqnarray}

The Lagrangian and the gauge transformation
for the massive bosonic higher spin field interacting with constant electromagnetic field
in the quartet formulation
are\footnote{
In order to obtain triplet formulation \cite{Francia:2002pt} one should to discard field $|s_4\rangle$ and Lagrangian multipliers $|a_2\rangle$, $|a_3\rangle$ in (\ref{Lagr-q}).}
\begin{eqnarray}
{\cal{}L}&=&
\langle{}s_1|
L_0|s_1\rangle-L_1^+|a_1\rangle-L_2^{+\prime}|a_2\rangle
\bigr\}
-
\langle{}s_2|
\bigl\{
L_0|s_2\rangle-L_1|a_1\rangle+|a_2\rangle-L_2^{+\prime}|a_3\rangle
\bigr\}
\nonumber
\\
&&
+
\langle{}s_4|
\Bigl\{
L_1|a_2\rangle-L_1^+|a_3\rangle
\Bigr\}
-
\langle{}a_1|
\Bigl\{
L_1|s_1\rangle-L_1^+|s_2\rangle -|a_1\rangle
\Bigr\}
\nonumber
\\[0.3em]
&&
-
\langle{}a_2|
\bigl\{
L_2'|s_1\rangle+|s_2\rangle -L_1^+|s_4\rangle
\bigr\}
+
\langle{}a_3|
\bigl\{
L_2'|s_2\rangle-L_1|s_4\rangle
\bigr\}
\label{Lagr-q}
\end{eqnarray}
\begin{align}
&\delta|s_1\rangle=L_1^+|\varepsilon\rangle
&&\delta|s_2\rangle= L_1|\varepsilon\rangle
&&\delta|s_4\rangle= L_2'|\varepsilon\rangle
\label{GT-q}
\end{align}
In the above relations the small letters in the state vectors
$|s_i\rangle$ and $|a_i\rangle$ denote the parts of the
corresponding vectors $|S_i\rangle$ and $|A_i\rangle$ depending only
on the oscillators
 $(a_\mu,b_1)$.
Also the operators $L_2'$ and $L_2^{+\prime}$ are the $(a_\mu,b_1)$ parts of the operators $L_2$ and $L_2^+$ (\ref{op-l2})
\begin{align}
&
L_2'=\tfrac{1}{2}a^{\mu}a_\mu+\tfrac{1}{2}b_1^2,
&&
L_2^{+\prime}=\tfrac{1}{2}a^{+\mu}a_\mu^++\tfrac{1}{2}b_1^{+2}
.
\end{align}

Next we will show that the Lagrangian formulation, which obtained in
\cite{Klishevich:1998ng}, is a particular case of our general result
(\ref{LagrF}). To get such Lagrangian formulations from (\ref{Lagr-q}) we first partly
fix the gauge, removing the field $|s_4\rangle$ with the help of
gauge transformations (\ref{GT-q}) and then integrate out all the
fields except the field $|s_1\rangle$. The result is
\begin{eqnarray}
{\cal{}L}&=&
\langle{}s_1|\Bigl\{L_0-2L_2^{+\prime}L_0L_2'-L_1^+L_1-L_1^+L_1^+L_2'-L_2^{+\prime}L_1L_1-L_2^{+\prime}L_1^+L_1L_2'\Bigr\}|s_1\rangle
.
\label{Lagr-Fron}
\\*[0.5em]
&&
\qquad\delta|s_1\rangle=L_1^+|\lambda_1\rangle,
\end{eqnarray}
where the state
 $|s_1\rangle$ and the parameter of gauge transformations $|\varepsilon\rangle$ obey the constraints
\begin{align}
&
L_2'L_2'|s_1\rangle=0,
&&
L_2'|\varepsilon\rangle=0.
\label{triple}
\end{align}
Such a partial form of the Lagrangian was obtained in
\cite{Klishevich:1998ng}, but with less general\footnote{It should
be noted that in \cite{Klishevich:1998ng} was considered deformation
of the operators corresponding to the traceless conditions as well.
But this deformation is proportional to an arbitrary constants and
as we said at the beginning of our paper can be removed by a field
redefinition.} expressions for the operators
(\ref{T1})--(\ref{op-L1+}).

We can proceed to obtain the different forms of the Lagrangian
formulation possessing the interesting properties. For example, we
can resolve constraints on the field and the gauge parameter
(\ref{triple}). Using decomposition (\ref{decomp-q}) for
$|\varepsilon\rangle$ and analogous decomposition for $|s_1\rangle$
\begin{eqnarray}
\label{res-psi}
|s_1\rangle=\sum_{k=0}^{s}\frac{1}{k!}(b_1^+)^{k}|\varphi_{s-k}\rangle
&\qquad&
|\varphi_k\rangle\propto\varphi_{\mu_1\cdots\mu_k}(x)a^{\mu_1+}\cdots a^{\mu_k+}|0\rangle\end{eqnarray}
we find that gauge parameters $|\varepsilon_{s-1}\rangle$ and $|\varepsilon_{s-2}\rangle$ are not restricted and the other parameters $|\epsilon_{k}\rangle$ are expressed in terms of their traces
\begin{eqnarray}
|\varepsilon_{s-2k-1}\rangle=(2l_2')^k|\varepsilon_{s-1}\rangle
&\qquad&
|\varepsilon_{s-2k-2}\rangle=(2l_2')^k|\varepsilon_{s-2}\rangle
\end{eqnarray}
where $l_2'$ was defined in (\ref{properators}).
So we may make gauge transformation using the unrestricted gauge parameters
$|\varepsilon_{s-1}\rangle$ and $|\varepsilon_{s-2}\rangle.$

One can do the same  for the field  $|s_1\rangle$. Due to restriction (\ref{triple}) there are only four independent fields
$|\varphi_{s}\rangle$, $|\varphi_{s-1}\rangle$, $|\varphi_{s-2}\rangle$, $|\varphi_{s-3}\rangle$
and all the other fields are expressed through these four fields
\begin{eqnarray}
|\varphi_{s-2k}\rangle&=&k(2l_2')^{k-1}|\varphi_{s-2}\rangle-(k-1)(2l_2')^k|\varphi_{s}\rangle
\\
|\varphi_{s-2k-1}\rangle&=&k(2l_2')^{k-1}|\varphi_{s-3}\rangle-(k-1)(2l_2')^k|\varphi_{s-1}\rangle
\end{eqnarray}

Thus one can obtain\footnote{Since the Lagrangian formulation is
very large, we do not present it here.} the gauge invariant
Lagrangian formulations for a  massive bosonic field interacting
with constant electromagnetic field with the help of four fields
$|\varphi_{s}\rangle$, $|\varphi_{s-1}\rangle$,
$|\varphi_{s-2}\rangle$, $|\varphi_{s-3}\rangle$ and two gauge
parameters $|\varepsilon_{s-1}\rangle$ and
$|\varepsilon_{s-2}\rangle$.

Finally,
using the remaining unrestricted gauge parameters
$|\varepsilon_{s-1}\rangle$ and $|\varepsilon_{s-2}\rangle$
one can remove fields $|\varphi_{s-1}\rangle$, $|\varphi_{s-2}\rangle$
using gauge transformation
and obtain the Lagrangian formulation in terms of two traceful
unrestricted fields: a field $|\varphi_{s}\rangle$ whose traceless
part is a physical field
 and an additional $|\varphi_{s-3}\rangle$ field.
This Lagrangian  has no gauge invariance since we have already used
entire gauge freedom. It should be noted that if we decompose the
traceful fields $|\varphi_{s}\rangle$ and $|\varphi_{s-3}\rangle$ in
a series of traceless fields we obtain set of the fields which
coincide with the set of fields of Singh and Hagen
\cite{Singh:1974qz}.

\section{Conclusions}
In the present paper we have developed the BRST approach to
 Lagrangian description of massive higher
spin bosonic fields coupled to a constant external electromagnetic
field and constructed the corresponding quartic interaction vertex.
Such a vertex is quadratic in dynamical higher spin fields and
quadratic in external field. To this end, we deformed the operators
underlying the BRST charge, which correspond to the noninteracting
bosonic massive higher spin fields, by the terms depending on the
electromagnetic field. Our main result is Lagrangian (\ref{LagrF})
which contains apart from the basic field also some number of
auxiliary fields which provide the gauge invariant description for
massive theory, and the  number of these fields grows with the value
of the spin.

We have shown that one can  partially or completely fix the gauge
invariance and obtain from the general Lagrangian (\ref{LagrF}) a
family of  different Lagrangian formulations with a smaller number
of auxiliary fields. As an example we have derived a Lagrangian
formulation for the massive bosonic higher spin fields coupled to a
constant electromagnetic background in the quartet (unconstrained)
formulation (\ref{Lagr-q})
\cite{Buchbinder:2007ak,Buchbinder:2008ss} and obtained the results
of paper \cite{Klishevich:1998ng} (\ref{Lagr-Fron}) as a particular case of the
general Lagrangian  (\ref{LagrF}).

Since in our previous paper \cite{Buchbinder:2015uea} we have
considered fermionic higher spin fields it would be naturally
interesting  to generalize the present results for the case of
supersymmetric systems\footnote{Lagrangian formulation of free
supersymmetric massive higher spin theory was done in \cite{Z}.} as
well as to consider higher order interactions. Inclusion of a
nontrivial gravitational background is another interesting problem
to consider (see for example \cite{Florakis:2014kfa} for recent
progress in these directions). It would be interesting also to
establish more connection of BRST approach with the recent studies
in conformal higher spin fields (see for example
\cite{Metsaev:2009ym}). We hope to address these questions in future
publications.

\section*{Acknowledgments}
I.L.B would like to acknowledge the Mainz Institute for Theoretical
Physics (MITP) for enabling me to complete some portion of this
work. I.L.B is grateful to the grant for LRSS, project
88.2014.2 and RFBR grants, projects No 13-02-90430 and No
15-02-03594 for partial support. Research of V.A.K was also
supported in part by Russian Ministry of Education and Science under
contract 3.867.2014/K.

\setcounter{equation}0
\appendix
\numberwithin{equation}{section}

\section{The arbitrary parameters in the first order in $F_{\mu\nu}$}
Below we give the expressions for free parameters which are present in the equations (\ref{op-L1})--(\ref{op-L1+})
\begin{align*}
&\alpha_{0(0)}=c_1+ic_2&&\alpha_{0(0)}'=c_1-ic_2
&&
\alpha_{0(k\ge1)}=-\alpha_{0(k\ge1)}'=\tfrac{(-2)^k}{k!}\,ic_2
\\&
\beta_{0(0)}=c_3+ic_4&&\beta_{0(0)}'=-c_3+ic_4
&&
\beta_{0(1)}=3-3c_1-4c_3+ic_2
\\&&&&&
\beta_{0(1)}'=-3+3c_1+4c_3+ic_2
\\&&&&&
\beta_{0(k\ge2)}=\tfrac{(-2)^{k-1}}{k!}[(k+1)(1-c_1-2c_3)+ic_2]
\\&&&&&
\beta_{0(k\ge2)}'=\tfrac{(-2)^{k-1}}{k!}[-(k+1)(1-c_1-2c_3)+ic_2]
\\&
\gamma_{0(0)}=-2c_3&&\gamma_{0(0)}'=2c_3
&&
\gamma_{0(k\ge1)}=-\gamma_{0(k\ge1)}'=\tfrac{(-2)^k}{k!}(1-c_1-2c_3)
\\
&
\xi_{0(0)}=1-2c_1
&&&&
\xi_{0(k\ge1)}=0
\\
&
\zeta_{0(0)}=3-2c_1
&&\zeta_{0(0)}'=-3+2c_1
&&
\zeta_{0((k\ge1)}=\zeta_{0((k\ge1)}'=0
\end{align*}
Here $c_1$, $c_2$, $c_3$, $c_4$ are arbitrary dimensionless constants.

\section{The arbitrary parameters in the second order in $F_{\mu\nu}$}
Below we give the expressions for free parameters which are present in the equations (\ref{op-L12})--(\ref{op-L1+2})
\begin{eqnarray*}
&&
\xi_{1(0)}=a_0
\qquad
\xi_{1(1)}=\tfrac{1}{2}(3-2c_1)
\qquad
\xi_{1(k\ge2)}=0
\\
&&
\xi_{2(0)}=\tfrac{1}{2}(3-2c_1)
\qquad
\xi_{2(1)}=3-2c_1
\qquad
\xi_{2(k\ge2)}=0
\\
&&
\xi_{3(0)}=\tfrac{11}{4}-7c_1+4c_1^2-4c_3+4c_1c_3+4c_3^2
\qquad
\xi_{3(k\ge1)}=0
\\
&&
\xi_{4(0)}=\xi_{4(0)}'=-\tfrac{1}{2}(3-2c_1)
\qquad
\xi_{4(k\ge1)}=\xi_{4(k\ge1)}'=0
\end{eqnarray*}
\vspace*{-2.5em}
\begin{eqnarray*}
&&
\zeta_{1(0)}=-4 c_1^2-6 c_3 c_1+8 c_1-4 c_3^2+7 c_3-\tfrac{17}{4}
+3 ic_4-2ic_1c_4
\\
&&
\zeta_{1(0)}'=-4 c_1^2-6 c_3 c_1+8c_1-4c_3^2+7c_3-\tfrac{17}{4}+2ic_4c_1-3ic_4
\\
&&
\zeta_{1(k\ge1)}=\tfrac{(-2)^{k-1}}{k!}(3-2c_1)(1-c_1-2c_3+ic_2)
\\
&&
\zeta_{1(k\ge1)}'=\tfrac{(-2)^{k-1}}{k!}(3-2c_1)(1-c_1-2c_3-ic_2)
\\
&&
\zeta_{2(0)}=6 c_1^2+8 c_3c_1-11 c_1+8 c_3^2-8 c_3+\tfrac{11}{2}-2 i c_2 c_1+3ic_2
\\
&&
\zeta_{2(0)}'=-6 c_1^2-8 c_3c_1+11 c_1-8 c_3^2+8 c_3-\tfrac{11}{2}-2 i c_2 c_1+3ic_2
\\
&&
\zeta_{2(k\ge1)}=\zeta_{2(k\ge1)}'=i\tfrac{(-2)^k}{k!}(3-2c_1)c_2
\\
&&
\zeta_{3(0)}=-c_3^2-c_4^2+2b_0
\qquad
\zeta_{3(1)}=-4 c_1^2-8 c_3 c_1+8 c_1-4c_3^2+10 c_3-\tfrac{17}{4}
\\
&&
\zeta_{3(k\ge2)}=-\tfrac{(-2)^{k-1}}{(k-1)!}(3-2c_1)(1-c_1-2c_3)
\\
&&
\zeta_{4(0)}=4 c_3^2+2 c_3+2c_1-\tfrac{7}{4}
\qquad
\zeta_{4(1)}=12 c_1^2+24 c_3 c_1-26 c_1+8c_3^2-32 c_3+\tfrac{29}{2}
\\
&&
\zeta_{4(k\ge2)}=-\tfrac{(-2)^k}{k!}(k+1)(3-2c_1)(1-c_1-2c_3)
\\
&&
\zeta_{5(0)}=\zeta_{5(0)}'=-\zeta_{4(0)}=\tfrac{7}{4}-4 c_3^2-2 c_3-2c_1
\\
&&
\zeta_{5(k\ge1)}=\zeta_{5(k\ge1)}'=\tfrac{(-2)^k}{k!}(3-2c_1)(1-c_1-2c_3)
\end{eqnarray*}
\vspace*{-2.5em}
\begin{eqnarray*}
&&
\alpha_{1(0)}=-\tfrac{1}{2}c_3-\tfrac{1}{2}a_0+ia_1
\qquad
\qquad
\alpha_{1(0)}'=-\tfrac{1}{2}c_3-\tfrac{1}{2}a_0-ia_1
\\
&&
\alpha_{1(1)}=c_1+c_3-\tfrac{9}{8}+i(-2a_1+a_2+3a_5)
\\
&&
\alpha_{1(1)}'=c_1+c_3-\tfrac{9}{8}-i(-2a_1+a_2+3a_5)
\\
&&
\alpha_{1(k\ge2)}=\tfrac{(-2)^{k-1}}{k!}\Bigl[\tfrac{1}{2}c_1+c_3-\tfrac{1}{2}
-i(2a_1-a_2-4a_5)+i(k-2)a_5\Bigr]
\\
&&
\alpha_{1(k\ge2)}'=\tfrac{(-2)^{k-1}}{k!}\Bigl[\tfrac{1}{2}c_1+c_3-\tfrac{1}{2}
+i(2a_1-a_2-4a_5)-i(k-2)a_5\Bigr]
\\
&&
\alpha_{2(0)}=\tfrac{1}{2}c_1-c_3-\tfrac{3}{4}+ia_2
\qquad
\qquad
\alpha_{2(0)}'=\tfrac{1}{2}c_1-c_3-\tfrac{3}{4}-ia_2
\\
&&
\alpha_{2(1)}=2c_1+2c_3-\frac{9}{4}+6ia_5
\qquad
\qquad
\alpha_{2(1)}'=2c_1+2c_3-\frac{9}{4}-6ia_5
\\
&&
\alpha_{2(k\ge2)}=\tfrac{(-2)^{k-1}}{k!}\Bigl[c_1+2c_3-1+2i(k+2)a_5\Bigr]
\\
&&
\alpha_{2(k\ge2)}'=\tfrac{(-2)^{k-1}}{k!}\Bigl[c_1+2c_3-1-2i(k+2)a_5\Bigr]
\\
&&
\alpha_{3(0)}=
-\tfrac{3}{2}c_1^2+\tfrac{7}{2}c_1-2c_1c_3+\tfrac{1}{2}c_2^2
-2c_3^2+2c_3-\tfrac{11}{8}+ia_3
\\
&&
\alpha_{3(0)}'=-\tfrac{3}{2}c_1^2+\tfrac{7}{2}c_1-2c_1c_3+\tfrac{1}{2}c_2^2
-2c_3^2+2c_3-\tfrac{11}{8}-ia_3
\\
&&
\alpha_{3(k\ge1)}=\tfrac{(-2)^k}{k!}ia_3
\qquad
\qquad
\alpha_{3(k\ge1)}'=-\tfrac{(-2)^k}{k!}ia_3
\\
&&
\alpha_{4(0)}=\tfrac{9}{8}-c_1-c_3-i(a_2+3a_5)
\qquad
\qquad
\alpha_{4(0)}'=\tfrac{9}{8}-c_1-c_3+i(a_2+3a_5)
\\
&&
\alpha_{4(1)}=\tfrac{1}{2}c_1+2 c_3-\frac{3}{8}+3ia_5
\qquad
\qquad
\alpha_{4(1)}'=\tfrac{1}{2}c_1+2 c_3-\frac{3}{8}-3ia_5
\\
&&
\alpha_{4(k\ge2)}=\tfrac{(-2)^{k-1}}{k!}\Bigl[c_1+2c_3-1+i(k+2)a_5\Bigr]
\\
&&
\alpha_{4(k\ge2)}'=\tfrac{(-2)^{k-1}}{k!}\Bigl[c_1+2c_3-1-i(k+2)a_5\Bigr]
\\
&&
\alpha_{5(0)}=\tfrac{5}{8}-\tfrac{1}{2}c_1+ia_5
\qquad
\qquad
\alpha_{5(0)}'=\tfrac{5}{8}-\tfrac{1}{2}c_1-ia_5
\\
&&
\alpha_{5(k\ge1)}=\tfrac{(-2)^k}{k!}ia_5
\qquad
\qquad
\alpha_{5(k\ge1)}'=-\tfrac{(-2)^k}{k!}ia_5
\end{eqnarray*}
\vspace*{-2.5em}
\begin{eqnarray*}
&&
\beta_{1(0)}=b_0+ib_1
\qquad
\qquad
\beta_{1(0)}'=b_0-ib_1
\\
&&
\beta_{1(1)}=
\tfrac{5}{2}c_1^2+5 c_3 c_1-5 c_1+\tfrac{1}{2}c_2^2+2 c_3^2-4 c_3+c_2 c_4+\tfrac{19}{8}
\\
&&\hspace{7ex}{}
+3 i c_4 c_1+4 i c_2 c_1-3 i c_2+3 i c_2 c_3-3 i c_4
+3ia_2+ia_3+6ia_5-4ib_1-2ib_2
\\
&&
\beta_{1(1)}'=\beta_{1(1)}^*
\\
&&
\beta_{1(2)}=
-\tfrac{5 }{2}c_1^2
-5 c_3 c_1
+4c_1-\tfrac{1}{2}c_2^2
+3 c_3-c_2c_4
-\tfrac{3}{2},
\\
&&\hspace{7ex}{}
-4 i c_2c_1-3 i c_4 c_1+3ic_2-3ic_2 c_3+3 i c_4
-3ia_2-ia_3-6ia_5+6ib_1+3ib_2
\\
&&
\beta_{1(2)}'=\beta_{1(2)}^*
\\
&&
\beta_{1((k\ge2)}=\tfrac{(-2)^{k-2}}{k!}\Bigl[2\beta_{1(2)}+(k-2)[\xi_{0(0)}\gamma_{0(1)}+\gamma_{1(1)}]\Bigr]
\qquad
\qquad
\beta_{1((k\ge2)}'=\beta_{1((k\ge2)}^*
\end{eqnarray*}

\vspace*{-2.0em}
\begin{eqnarray*}
&&
\beta_{2(0)}=
-\tfrac{7}{8}+c_1+c_3-2c_3^2+ib_2
\qquad\qquad
\beta_{2(0)}'=
\tfrac{7}{8}-c_1-c_3+2c_3^2+ib_2
\\
&&
\beta_{2(1)}=
-\tfrac{19}{4}-6c_1^2+11c_1-12c_1c_3+8c_3-4c_3^2
-2ic_1c_2-2ia_3
\\
&&
\beta_{2(1)}'=
\tfrac{19}{4}+6c_1^2-11c_1+12c_1c_3-8c_3+4c_3^2
-2ic_1c_2-2ia_3
\\
&&
\beta_{2(2)}=3(1-2c_1)(1-c_1-2c_3)+2ic_1c_2+2ia_3
\\
&&
\beta_{2(2)}'=-3(1-2c_1)(1-c_1-2c_3)+2ic_1c_2+2ia_3
\\
&&
\beta_{2((k\ge2)}=-\tfrac{(-2)^{k-1}}{k!}\Bigl[(k+1)(1-2c_1)(1-c_1-2c_3)+2ic_1c_2+2ia_3\Bigr]
\\
&&
\beta_{2((k\ge2)}'=\tfrac{(-2)^{k-1}}{k!}\Bigl[(k+1)(1-2c_1)(1-c_1-2c_3)-2ic_1c_2-2ia_3\Bigr]
\\
&&
\gamma_{1(0)}=\tfrac{7}{8}-c_1-c_3+2c_3^2-2ib_1-ib_2
\qquad
\qquad
\gamma_{1(0)}'=\tfrac{7}{8}-c_1-c_3+2c_3^2+2ib_1+ib_2
\\
&&
\gamma_{1(1)}=
 (1-2c_1)(1-c_1-2c_3)
\\
&&\hspace{7ex}{}
-2 i c_2 c_1-2 i c_4c_1+2 i c_2-2 i c_2c_3+2 i c_4
-2ia_2-4ia_5+4ib_1+2ib_2
\\
&&
\gamma_{1(1)}'=\gamma_{1(1)}^*
\qquad
\qquad
\gamma_{1(k\ge1)}=\tfrac{(-2)^{k-1}}{k!}\gamma_{1(1)}
\qquad
\qquad
\gamma_{1(k\ge1)}'=\tfrac{(-2)^{k-1}}{k!}\gamma_{1(1)}^*
\\
&&
\gamma_{2(0)}=-\gamma_{2(0)}'
=
\tfrac{7}{4}-2c_1-2c_3+4c_3^2
\qquad
\gamma_{2(k\ge1)}=-\gamma_{2(k\ge1)}'=-\tfrac{(-2)^k}{k!}(1-2c_1)(1-c_1-2c_3)
\end{eqnarray*}
Here
independent constants are
$c_1$, $c_2$, $c_4$,
$a_0$, $a_1$, $a_2$, $a_3$, $a_5$,
$b_0$,  $b_1$,  $b_2$
and
$c_3$ is expressed form equation
\begin{equation}
6 c_1^2+12 c_3 c_1-9 c_1+4 c_3^2-6 c_3+3=0
\label{c1c3}
\end{equation}
and has the form
$c_3=\frac{1}{4}\left(3-6 c_1\pm\sqrt{3}\sqrt{4 c_1^2-1}\right).$

\end{document}